\documentclass[a4paper,11pt]{article}
\usepackage{pos}
\usepackage{subcaption}

\newcommand*\diff{\mathop{}\!\mathrm{d}}

\newcommand{\ftop}{t\bar{t}t\bar{t}}

\title{NNLL resummation for the production of four top quarks}

\author[a]{Melissa van Beekveld}
\author[b]{Anna Kulesza}
\author[b]{Michele Lupattelli}
\author*[a]{Tommaso Saracco}

\affiliation[a]{Nikhef, Theory Group, Science Park 105, 1098 XG, Amsterdam, The Netherlands}
\affiliation[b]{Institute for Theoretical Physics, University of Münster,
Wilhelm-Klemm-Straße 9, \\ D-48149 Münster, Germany}

\emailAdd{mbeekvel@nikhef.nl}
\emailAdd{anna.kulesza@uni-muenster.de}
\emailAdd{michele.lupattelli@uni-muenster.de}
\emailAdd{tsaracco@nikhef.nl}

\abstract{Four-top production is one of the rarest processes of the Standard Model observable at the Large Hadron Collider, offering sensitivity to the top Yukawa coupling, Higgs width, and beyond-Standard-Model physics. We present precise predictions for the invariant-mass distribution and total cross section using threshold resummation at next-to-next-to-leading logarithmic (NNLL) accuracy, matched to next-to-leading order (NLO) calculations.}

\FullConference{14th Edition of the Large Hadron Collider Physics (LHCP2026)\\
18-22 May 2026\\
Paris, France\\}

\begin{document}
\maketitle

\section{Introduction} \label{sec:introduction}

One of the rarest processes ever measured at the LHC is the simultaneous production of four top quarks. Due to its small cross section and to the presence of four top quarks, it is a process of great physical interest: it provides sensitivity to the top Yukawa coupling~\cite{Cao:2016wib, Cao:2019ygh} and the Higgs boson width~\cite{ATLAS:2024mhs}. Furthermore, it enables the exploration of beyond the Standard Model scenarios~\cite{Darme:2018dvz,Toharia:2005gm,Craig:2016ygr,Dicus:1994bm,Farrar:1978xj,Beck:2015cga,Calvet:2012rk,Plehn:2008ae,Craig:2015jba, Abasov:2024mwk, Shooshtari:2026eyz}, and can be used to constrain operators in effective field theories~\cite{Hartland:2019bjb,Ethier:2021bye,Aoude:2022deh,Zhang:2017mls,Aguilar-Saavedra:2018ksv,Banelli:2020iau,Darme:2021gtt,Aleshko:2026kph}. 
The total cross-section for four-top production has been only recently measured by the ATLAS~\cite{ATLAS:2018kxv, ATLAS:2020hpj, ATLAS:2021kqb, ATLAS:2023ajo} and CMS~\cite{CMS:2019jsc, CMS:2019rvj, CMS:2023ftu} collaborations. The latest 
ATLAS measurement reads $22.5^{+6.6}_{-5.5}$~fb~\cite{ATLAS:2023ajo}, while the  CMS collaboration reports $17.7^{+4.4}_{-4.0}$~fb~\cite{CMS:2023ftu}. 

The most accurate fixed-order predictions are provided at NLO in both QCD and electroweak (EW) in Ref.~\cite{Frederix:2017wme}, with additional work on top decays in Refs.~\cite{Jezo:2021smh, Dimitrakopoulos:2024qib, Dimitrakopoulos:2024yjm}.
The inclusion of all-order effects via threshold resummation has been studied in Refs.~\cite{vanBeekveld:2022hty,vanBeekveld:2025ghw} at next-to-leading (NLL) accuracy, together with relative-order $\mathcal{O}(\alpha_s)$ non-logarithmic corrections (NLL$'$ accuracy).

In this contribution, we present preliminary results in the invariant-mass threshold resummation formalism at NNLL accuracy. The remainder of this work is organised as follows: in Sec.~\ref{sec:theory} we briefly review the theory behind this resummation approach, and summarise the ingredients needed to reach our level of accuracy; we then show results for the total cross section and the invariant-mass distribution in Sec.~\ref{sec:results}; we conclude with Sec.~\ref{sec:conclusions}, where we summarise our findings and outline next steps.

\section{Theoretical framework} \label{sec:theory}

The perturbative expansion in powers of $\alpha_s$ of the invariant-mass distribution $d\sigma/dQ$ contains logarithmically-enhanced contributions in the region where the partonic centre-of-mass energy $\sqrt{\hat{s}}$ is close to $Q$, where $Q$ is the invariant mass of the four final-state top quarks.
Specifically, these logarithms appear in plus distributions involving the partonic threshold variable $\hat\rho\equiv Q^2/\hat{s}$,
\begin{equation} \label{eq:logs}
    \alpha_s^n\; \left[\frac{\log^m(1-\hat{\rho})}{1-\hat{\rho}}\right]_+ \;,\quad m \leq 2n-1 \;.
\end{equation}
The $\hat{\rho}\to1$ region corresponds to the configuration where almost all of the energy available to the process is used to produce the four top quarks, hence any additional radiation must be soft. In this limit, the soft and collinear contributions factorise from the hard scattering process.
This allows to resum the logarithms of Eq.~\eqref{eq:logs} to all orders.
However, the factorisation of the phase space is only achieved in a conjugate space. We thus work in Mellin space, performing a Mellin transform with respect to the hadronic threshold variable $\rho = x_1 x_2 \hat{\rho}$,
\begin{equation}
    \frac{\diff\tilde{\sigma}(N)}{\diff Q}=\int_0^1 \diff\rho\, \rho^{N-1}\,\frac{\diff\sigma(\rho)}{\diff Q}\;.
\end{equation}
The Mellin transform maps the logarithms of Eq.~\eqref{eq:logs} into logarithms of the Mellin moment $N$, specifically to powers of $\log \bar{N}$, where  $\bar{N}\equiv N e^{\gamma_E}$.
The hadronic differential cross section in Mellin space can then be written as
\begin{equation} \label{eq:xsec_factorisation}
    \frac{\diff\tilde{\sigma}^{\text{res}}(N)}{\diff Q} =  \tilde{f}_1(N+1)\tilde{f}_2(N+1)\;\text{Tr}\left[\textbf{S}(N+1)\; \textbf{H}\right]\,\Delta_1(N+1)\Delta_2(N+1)\;.
\end{equation}
The functions $\tilde{f}_i$ are the Mellin transforms of the parton distribution functions (PDFs).
The underlying hard process is described by the hard function $\textbf{H}=\textbf{H}^{(0)}+\frac{\alpha_s}{4\pi}\,\textbf{H}^{(1)}+\mathcal{O}(\alpha_s^2)$. At NNLL accuracy, it is needed at one loop, and $\textbf{H}^{(1)}$  can be written as $\textbf{H}^{(1)}=\textbf{V}^{(1)}+\textbf{C}^{(1)}$,
where $\textbf{V}^{(1)}$ are the one-loop virtual corrections, extracted from OpenLoops \cite{Ossola:2007ax,vanHameren:2009dr,vanHameren:2010cp,Cascioli:2011va,Denner:2016kdg,Buccioni:2017yxi,Buccioni:2019sur} , and $\textbf{C}^{(1)}$ accounts for relative order $\mathcal{O}(\alpha_s)$ (soft-) collinear $N$-independent contributions.
The hard function is a matrix in colour space, and multiplies the collinear-subtracted soft function matrix $\textbf{S}$, which accounts for wide-angle soft-gluon emissions.
The soft function is the solution of a (matrix) renormalisation group equation and can be written as $\textbf{S}=\bar{\textbf{U}}\,\tilde{\textbf{S}}\,\textbf{U}$, where the matrix $\tilde{\textbf{S}}$ is the boundary condition, needed at one loop, and $\textbf{U}$ is the evolution matrix. The latter can be written as a path-ordered exponential
\begin{equation} \label{eq:Ufunction}
     \textbf{U}(N)=\mathcal{P}\exp\left[\frac12 \int_{\mu_R^2}^{Q^2/\bar{N}^2} \frac{\diff \mu^2}{\mu^2} \,{\bf\Gamma}(\mu^2,\alpha_s(\mu^2)) \right]\;,
\end{equation}
where ${\bf\Gamma}=  \frac{\alpha_s}{4\pi}\, {\bf\Gamma}^{(1)} + \left(\frac{\alpha_s}{4\pi}\right)^2 {\bf\Gamma}^{(2)}+\mathcal{O}(\alpha_s^3)$ is the soft anomalous dimension matrix. At NLL and NLL$'$ only the one-loop contribution is needed, but at NNLL accuracy also the two-loop term is required. The one- and two-loop coefficients can be calculated perturbatively from the UV poles of eikonal contributions, and their analytical expressions can be found in Refs.~\cite{Ferroglia:2009ep,Ferroglia:2009ii}. 
Since the soft anomalous dimension depends on the kinematics of the process, in order to reduce Eq.~\eqref{eq:Ufunction} to an ordinary exponential, it needs to be diagonalised for every phase-space point. However, it is not in general possible to diagonalise the one- and two-loop coefficients simultaneously, so the technique outlined in Refs.~\cite{Buras:1979yt, Ahrens:2010zv, Kulesza:2017ukk} is used. 
The soft anomalous dimension develops Coulomb-enhanced terms when a subset of the heavy particles has vanishing relative velocity~\cite{Ferroglia:2009ep, Ferroglia:2009ii,Czakon:2009zw,Beneke:2009rj}. Since Coulomb ladders are not resummed here, the Coulomb contributions are retained through the NLO hard function, while the appropriate zero-relative-velocity limit is taken in the soft anomalous dimension \cite{Czakon:2009zw,Beneke:2009rj,Abreu:2022cco}; this generalises the one-loop treatment of Ref~\cite{vanBeekveld:2025ghw}. 

The remaining contributions in Eq.~\eqref{eq:xsec_factorisation} that need to be discussed are the functions $\Delta_i$, which contain the (soft-)collinear radiation from the incoming particles. They can be written as 
\begin{equation} \label{eq:jetfunction}
    \Delta_i(N)=\exp \left\{\frac{1}{\alpha_s} g_1(\lambda) + g_2(\lambda) + \alpha_s g_3(\lambda) + \mathcal{O}(\alpha_s^2)
    \right\}\;,
\end{equation}
where $\lambda\equiv\alpha_s b_0 \log\bar{N}$. At NNLL accuracy, we need the $g_1$, $g_2$ and $g_3$ functions. Their analytical expressions can be found, e.g., in Refs.~\cite{Catani:2003zt, vanBeekveld:2019cks}.

Finally, to obtain physical results, we perform an inverse Mellin transform to return to momentum space. The inversion is performed numerically using the minimal prescription method~\cite{Catani:1996yz}.
Finally, we obtain the prediction at NLO+NNLL accuracy by matching our resummed result to the fixed-order at NLO:
\begin{equation} \label{eq:matching}
    \diff \sigma^{\text{NLO}+\text{NNLL}}=\diff\sigma^{\text{NLO}} + \left[\diff\sigma^{\text{NNLL}}-\diff\sigma^{\text{NNLL}}|_{\text{NLO}}\right],
\end{equation}
where the last term indicates the NNLL-resummed cross section expanded up to NLO, necessary to avoid the double counting of relative-order $\mathcal{O}(\alpha_s)$ terms.

\section{Results} \label{sec:results}
We present preliminary results for the invariant-mass distribution $d\sigma/dQ$ and the
total cross section at $\sqrt{S}=13.6~\text{TeV}$. The resummed predictions are
matched to fixed-order NLO in QCD+EW\footnote{To be precise, we consider QCD contributions of $\mathcal{O}(\alpha_s^4)$ and $\mathcal{O}(\alpha_s^5)$, and the dominant EW corrections~\cite{Frederix:2017wme}, i.e.~$\mathcal{O}(\alpha_s^3\alpha)$, $\mathcal{O}(\alpha_s^2\alpha^2)$,
$\mathcal{O}(\alpha_s^4\alpha)$ and $\mathcal{O}(\alpha_s^3\alpha^2)$, and label this accuracy simply as ``NLO'' from now on.} computed with
MG5\_aMC@NLO v3.5.5 \cite{Alwall:2014hca, Frederix:2018nkq}. The central scale is set
to $\mu_0 = Q$, where $Q$ is the invariant mass of the four-top system. Theoretical uncertainties are
estimated via 7-point scale variation.
We use the PDF set LUXqed\_plus\_PDF4LHC15\_nnlo\_100~\cite{Manohar:2016nzj,Manohar:2017eqh,Butterworth:2015oua,NNPDF:2014otw,Harland-Lang:2014zoa,Dulat:2015mca}, which includes the photon content of the proton.

\begin{figure}[t]
\centering
\includegraphics[width=0.55\textwidth]{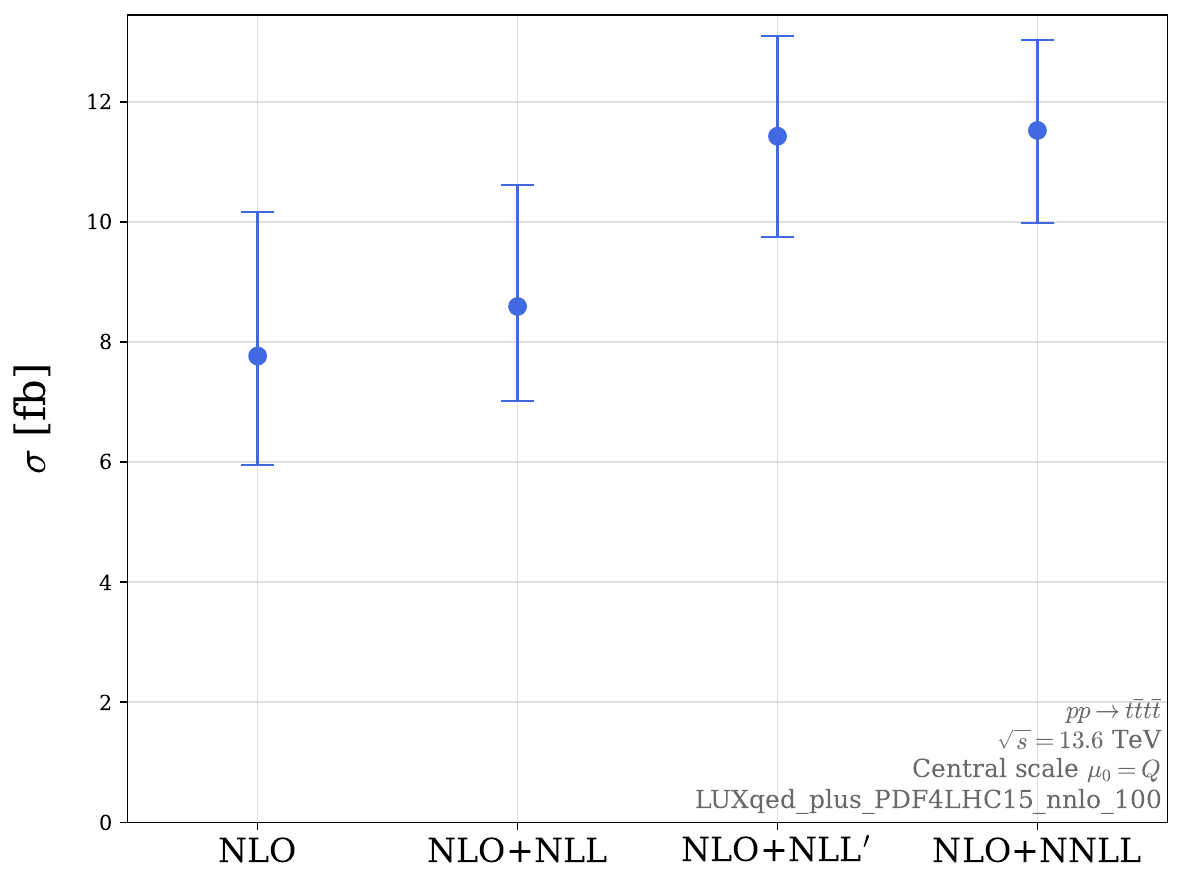}
\caption{Total cross section for $pp \to t\bar{t}t\bar{t}$ at $\sqrt{s}=13.6~\text{TeV}$
for increasing levels of accuracy (NLO, NLO+NLL, NLO+NLL$'$, NLO+NNLL).
Error bars represent the uncertainty from 7-point scale variation.}
\label{fig:xsec}
\end{figure}

In Fig.~\ref{fig:xsec} we show the total cross section at increasing levels of accuracy, showing the fixed-order NLO result and the matched resummed calculations at NLO+NLL, NLO+NLL$'$, and NLO+NNLL.
We note that for the scale $\mu_0=Q$ an increase in the central value is observed as higher logarithmic orders are included. The cross section grows by $48\%$ going from the NLO prediction to the NLO+NNLL resummed result. 
We also observe a progressive reduction of the scale uncertainty.
Furthermore, we note that the predictions stabilise between NLO+NLL$'$ and NLO+NNLL, with a relative increase in the central value, and a relative reduction of the theoretical errors, both at the percent level.

\begin{figure}[t]
\centering
\includegraphics[width=0.62\textwidth]{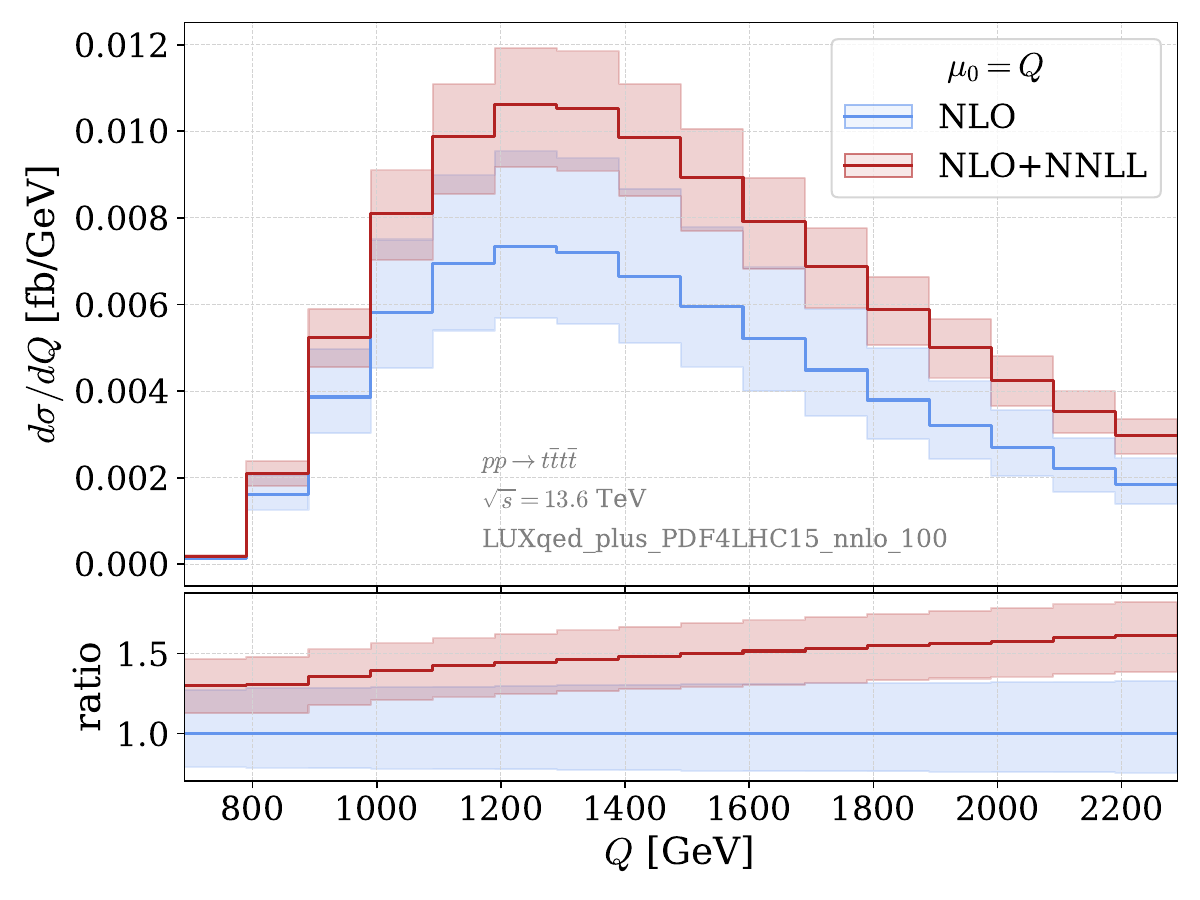}
\caption{Invariant-mass distribution of the $t\bar{t}t\bar{t}$ system at NLO
(blue) and NLO+NNLL (red) accuracy. Error bands represent the theoretical uncertainty estimated through 7-point scale variation.
The lower panel shows the ratio to the NLO prediction. }
\label{fig:invmassdistr}
\end{figure}

The invariant-mass distribution $d\sigma/dQ$ is shown in Fig.~\ref{fig:invmassdistr}. We show the results for NLO
(blue) and NLO+NNLL (red), with bands indicating the uncertainty from 7-point scale variation.
The resummed result at NLO+NNLL accuracy exhibits a reduction of the scale-variation uncertainty across the entire invariant-mass spectrum. It shows a significant shape modification relative to the NLO distribution, with a relative correction of $+30\%$ in the first bin, increasing with higher values of the invariant mass, away from the four-top absolute threshold region, and reaching $+61\%$ at the end of the invariant-mass range considered.

\section{Conclusions} \label{sec:conclusions}

In this contribution we have presented preliminary predictions for $pp\to\ftop$ at NLO+NNLL accuracy, which represent an improvement with respect to the previous NLO+NLL$'$ calculation~\cite{vanBeekveld:2025ghw}. 
For the central scale choice $\mu_0=Q$, we notice a moderate increase in the value of the cross section at NLO+NNLL compared to NLO+NLL$'$, suggesting an improved stability of the logarithmic expansion for the observables shown.
The next step is to provide a more detailed analysis of the theoretical uncertainties entering the calculation, including scale variations, PDFs and $\alpha_s$ uncertainties, top-mass dependence, and the impact of subleading contributions in the threshold resummation framework.


\acknowledgments{
We thank Federico Buccioni for providing the  colour-decomposed one-loop amplitudes in the \textsf{OpenLoops} software package. 
MvB is supported by the Dutch Research Council (NWO) under  project number VI.Veni.232.190. 
Part of the calculations for this publication were performed on the HPC cluster PALMA II of the University of Münster, subsidised by the DFG (INST 211/667-1). TS thanks the University of Oxford and the Particle Theory group for the hospitality while part of this contribution was prepared. 
}

\bibliographystyle{JHEP.bst}
\bibliography{References}

@article{ATLAS:2018kxv,
    author = "Aaboud, Morad and others",
    collaboration = "ATLAS",
    title = "{Search for four-top-quark production in the single-lepton and opposite-sign dilepton final states in pp collisions at $\sqrt{s}$ = 13 TeV with the ATLAS detector}",
    eprint = "1811.02305",
    archivePrefix = "arXiv",
    primaryClass = "hep-ex",
    reportNumber = "CERN-EP-2018-174",
    doi = "10.1103/PhysRevD.99.052009",
    journal = "Phys. Rev. D",
    volume = "99",
    number = "5",
    pages = "052009",
    year = "2019"
}

@article{ATLAS:2020hpj,
    author = "Aad, Georges and others",
    collaboration = "ATLAS",
    title = "{Evidence for $t\bar{t}t\bar{t}$ production in the multilepton final state in proton\textendash{}proton collisions at $\sqrt{s}=13$ $\text {TeV}$ with the ATLAS detector}",
    eprint = "2007.14858",
    archivePrefix = "arXiv",
    primaryClass = "hep-ex",
    reportNumber = "CERN-EP-2020-111",
    doi = "10.1140/epjc/s10052-020-08509-3",
    journal = "Eur. Phys. J. C",
    volume = "80",
    number = "11",
    pages = "1085",
    year = "2020"
}

@article{ATLAS:2021kqb,
    author = "Aad, Georges and others",
    collaboration = "ATLAS",
    title = "{Measurement of the t$ \overline{t} $t$ \overline{t} $ production cross section in $pp$ collisions at $ \sqrt{s} $ = 13 TeV with the ATLAS detector}",
    eprint = "2106.11683",
    archivePrefix = "arXiv",
    primaryClass = "hep-ex",
    reportNumber = "CERN-EP-2021-075",
    doi = "10.1007/JHEP11(2021)118",
    journal = "JHEP",
    volume = "11",
    pages = "118",
    year = "2021"
}

@article{ATLAS:2023ajo,
    author = "Aad, Georges and others",
    collaboration = "ATLAS",
    title = "{Observation of four-top-quark production in the multilepton final state with the ATLAS detector}",
    eprint = "2303.15061",
    archivePrefix = "arXiv",
    primaryClass = "hep-ex",
    reportNumber = "CERN-EP-2023-055",
    doi = "10.1140/epjc/s10052-023-11573-0",
    journal = "Eur. Phys. J. C",
    volume = "83",
    number = "6",
    pages = "496",
    year = "2023",
    note = "[Erratum: Eur.Phys.J.C 84, 156 (2024)]"
}

@article{CMS:2019rvj,
    author = "Sirunyan, Albert M and others",
    collaboration = "CMS",
    title = "{Search for production of four top quarks in final states with same-sign or multiple leptons in proton-proton collisions at $\sqrt{s}=$ 13 TeV}",
    eprint = "1908.06463",
    archivePrefix = "arXiv",
    primaryClass = "hep-ex",
    reportNumber = "CMS-TOP-18-003, CERN-EP-2019-163",
    doi = "10.1140/epjc/s10052-019-7593-7",
    journal = "Eur. Phys. J. C",
    volume = "80",
    number = "2",
    pages = "75",
    year = "2020"
}

@article{CMS:2019jsc,
    author = "Sirunyan, Albert M and others",
    collaboration = "CMS",
    title = "{Search for the production of four top quarks in the single-lepton and opposite-sign dilepton final states in proton-proton collisions at $ \sqrt{s} $ = 13 TeV}",
    eprint = "1906.02805",
    archivePrefix = "arXiv",
    primaryClass = "hep-ex",
    reportNumber = "CERN-EP-2019-098",
    doi = "10.1007/JHEP11(2019)082",
    journal = "JHEP",
    volume = "11",
    pages = "082",
    year = "2019"
}

@article{CMS:2023ftu,
    author = "Hayrapetyan, Aram and others",
    collaboration = "CMS",
    title = "{Observation of four top quark production in proton-proton collisions at s=13TeV}",
    eprint = "2305.13439",
    archivePrefix = "arXiv",
    primaryClass = "hep-ex",
    reportNumber = "CMS-TOP-22-013, CERN-EP-2023-090",
    doi = "10.1016/j.physletb.2023.138290",
    journal = "Phys. Lett. B",
    volume = "847",
    pages = "138290",
    year = "2023"
}

@article{Cao:2016wib,
    author = "Cao, Qing-Hong and Chen, Shao-Long and Liu, Yandong",
    title = "{Probing Higgs Width and Top Quark Yukawa Coupling from $t\bar{t}H$ and $t\bar{t}t\bar{t}$ Productions}",
    eprint = "1602.01934",
    archivePrefix = "arXiv",
    primaryClass = "hep-ph",
    doi = "10.1103/PhysRevD.95.053004",
    journal = "Phys. Rev. D",
    volume = "95",
    number = "5",
    pages = "053004",
    year = "2017"
}

@article{Cao:2019ygh,
    author = "Cao, Qing-Hong and Chen, Shao-Long and Liu, Yandong and Zhang, Rui and Zhang, Ya",
    title = "{Limiting top quark-Higgs boson interaction and Higgs-boson width from multitop productions}",
    eprint = "1901.04567",
    archivePrefix = "arXiv",
    primaryClass = "hep-ph",
    doi = "10.1103/PhysRevD.99.113003",
    journal = "Phys. Rev. D",
    volume = "99",
    number = "11",
    pages = "113003",
    year = "2019"
}

@article{Darme:2018dvz,
    author = "Darm\'e, Luc and Fuks, Benjamin and Goodsell, Mark",
    title = "{Cornering sgluons with four-top-quark events}",
    eprint = "1805.10835",
    archivePrefix = "arXiv",
    primaryClass = "hep-ph",
    doi = "10.1016/j.physletb.2018.08.001",
    journal = "Phys. Lett. B",
    volume = "784",
    pages = "223--228",
    year = "2018"
}

@article{Toharia:2005gm,
    author = "Toharia, Manuel and Wells, James D.",
    title = "{Gluino decays with heavier scalar superpartners}",
    eprint = "hep-ph/0503175",
    archivePrefix = "arXiv",
    reportNumber = "MCTP-05-46",
    doi = "10.1088/1126-6708/2006/02/015",
    journal = "JHEP",
    volume = "02",
    pages = "015",
    year = "2006"
}

@article{Craig:2016ygr,
    author = "Craig, Nathaniel and Hajer, Jan and Li, Ying-Ying and Liu, Tao and Zhang, Hao",
    title = "{Heavy Higgs bosons at low $\tan \beta$: from the LHC to 100 TeV}",
    eprint = "1605.08744",
    archivePrefix = "arXiv",
    primaryClass = "hep-ph",
    doi = "10.1007/JHEP01(2017)018",
    journal = "JHEP",
    volume = "01",
    pages = "018",
    year = "2017"
}

@article{Dicus:1994bm,
    author = "Dicus, D. and Stange, A. and Willenbrock, S.",
    title = "{Higgs decay to top quarks at hadron colliders}",
    eprint = "hep-ph/9404359",
    archivePrefix = "arXiv",
    reportNumber = "CPP-94-18, BNL-60339, ILL-TH-94-9",
    doi = "10.1016/0370-2693(94)91017-0",
    journal = "Phys. Lett. B",
    volume = "333",
    pages = "126--131",
    year = "1994"
}

@article{Farrar:1978xj,
    author = "Farrar, Glennys R. and Fayet, Pierre",
    title = "{Phenomenology of the Production, Decay, and Detection of New Hadronic States Associated with Supersymmetry}",
    reportNumber = "CALT-68-648",
    doi = "10.1016/0370-2693(78)90858-4",
    journal = "Phys. Lett. B",
    volume = "76",
    pages = "575--579",
    year = "1978"
}

@article{Beck:2015cga,
    author = "Beck, Lana and Blekman, Freya and Dobur, Didar and Fuks, Benjamin and Keaveney, James and Mawatari, Kentarou",
    title = "{Probing top-philic sgluons with LHC Run I data}",
    eprint = "1501.07580",
    archivePrefix = "arXiv",
    primaryClass = "hep-ph",
    doi = "10.1016/j.physletb.2015.04.043",
    journal = "Phys. Lett. B",
    volume = "746",
    pages = "48--52",
    year = "2015"
}

@article{Calvet:2012rk,
    author = "Calvet, Samuel and Fuks, Benjamin and Gris, Philippe and Valery, Loic",
    title = "{Searching for sgluons in multitop events at a center-of-mass energy of 8 TeV}",
    eprint = "1212.3360",
    archivePrefix = "arXiv",
    primaryClass = "hep-ph",
    reportNumber = "CERN-PH-TH-2012-353, IPHC-PHENO-12-08, PCCF-R1-12-08",
    doi = "10.1007/JHEP04(2013)043",
    journal = "JHEP",
    volume = "04",
    pages = "043",
    year = "2013"
}

@article{Plehn:2008ae,
    author = "Plehn, Tilman and Tait, Tim M. P.",
    title = "{Seeking Sgluons}",
    eprint = "0810.3919",
    archivePrefix = "arXiv",
    primaryClass = "hep-ph",
    reportNumber = "ANL-HEP-PR-08-65, EDINBURGH-2008-43, NU-HEP-TH-08-08",
    doi = "10.1088/0954-3899/36/7/075001",
    journal = "J. Phys. G",
    volume = "36",
    pages = "075001",
    year = "2009"
}

@article{Craig:2015jba,
    author = "Craig, Nathaniel and D'Eramo, Francesco and Draper, Patrick and Thomas, Scott and Zhang, Hao",
    title = "{The Hunt for the Rest of the Higgs Bosons}",
    eprint = "1504.04630",
    archivePrefix = "arXiv",
    primaryClass = "hep-ph",
    doi = "10.1007/JHEP06(2015)137",
    journal = "JHEP",
    volume = "06",
    pages = "137",
    year = "2015"
}

@article{Abasov:2024mwk,
    author = "Abasov, E. and others",
    title = "{Search for dark matter mediator in the production of three and four top quarks}",
    eprint = "2407.08308",
    archivePrefix = "arXiv",
    primaryClass = "hep-ph",
    month = "7",
    year = "2024"
}

@article{Hartland:2019bjb,
    author = "Hartland, Nathan P. and Maltoni, Fabio and Nocera, Emanuele R. and Rojo, Juan and Slade, Emma and Vryonidou, Eleni and Zhang, Cen",
    title = "{A Monte Carlo global analysis of the Standard Model Effective Field Theory: the top quark sector}",
    eprint = "1901.05965",
    archivePrefix = "arXiv",
    primaryClass = "hep-ph",
    reportNumber = "OUTP-18-07P, Nikhef-2018-058, CP3-19-02, CERN-TH-2018-274",
    doi = "10.1007/JHEP04(2019)100",
    journal = "JHEP",
    volume = "04",
    pages = "100",
    year = "2019"
}

@article{Ethier:2021bye,
    author = "Ethier, Jacob J. and Magni, Giacomo and Maltoni, Fabio and Mantani, Luca and Nocera, Emanuele R. and Rojo, Juan and Slade, Emma and Vryonidou, Eleni and Zhang, Cen",
    collaboration = "SMEFiT",
    title = "{Combined SMEFT interpretation of Higgs, diboson, and top quark data from the LHC}",
    eprint = "2105.00006",
    archivePrefix = "arXiv",
    primaryClass = "hep-ph",
    reportNumber = "OUTP-20-05P, Nikhef-2020-020, CP3-21-12, MCNET-21-07,
  MAN/HEP/2021/004",
    doi = "10.1007/JHEP11(2021)089",
    journal = "JHEP",
    volume = "11",
    pages = "089",
    year = "2021"
}

@article{Aoude:2022deh,
    author = "Aoude, Rafael and El Faham, Hesham and Maltoni, Fabio and Vryonidou, Eleni",
    title = "{Complete SMEFT predictions for four top quark production at hadron colliders}",
    eprint = "2208.04962",
    archivePrefix = "arXiv",
    primaryClass = "hep-ph",
    reportNumber = "CP3-22-37",
    doi = "10.1007/JHEP10(2022)163",
    journal = "JHEP",
    volume = "10",
    pages = "163",
    year = "2022"
}

@article{Zhang:2017mls,
    author = "Zhang, Cen",
    title = "{Constraining $qqtt$ operators from four-top production: a case for enhanced EFT sensitivity}",
    eprint = "1708.05928",
    archivePrefix = "arXiv",
    primaryClass = "hep-ph",
    doi = "10.1088/1674-1137/42/2/023104",
    journal = "Chin. Phys. C",
    volume = "42",
    number = "2",
    pages = "023104",
    year = "2018"
}

@article{Aguilar-Saavedra:2018ksv,
    author = "Barducci, D. and others",
    editor = "Aguilar-Saavedra, Juan Antonio and Degrande, C. and Durieux, G. and Maltoni, F. and Vryonidou, E. and Zhang, C.",
    title = "{Interpreting top-quark LHC measurements in the standard-model effective field theory}",
    eprint = "1802.07237",
    archivePrefix = "arXiv",
    primaryClass = "hep-ph",
    reportNumber = "CERN-LPCC-2018-01",
    month = "2",
    year = "2018"
}

@article{Banelli:2020iau,
    author = "Banelli, Giovanni and Salvioni, Ennio and Serra, Javi and Theil, Tobias and Weiler, Andreas",
    title = "{The Present and Future of Four Top Operators}",
    eprint = "2010.05915",
    archivePrefix = "arXiv",
    primaryClass = "hep-ph",
    reportNumber = "TUM-HEP-1286-20, CERN-TH-2020-166",
    doi = "10.1007/JHEP02(2021)043",
    journal = "JHEP",
    volume = "02",
    pages = "043",
    year = "2021"
}

@article{Darme:2021gtt,
    author = "Darm\'e, Luc and Fuks, Benjamin and Maltoni, Fabio",
    title = "{Top-philic heavy resonances in four-top final states and their EFT interpretation}",
    eprint = "2104.09512",
    archivePrefix = "arXiv",
    primaryClass = "hep-ph",
    doi = "10.1007/JHEP09(2021)143",
    journal = "JHEP",
    volume = "09",
    pages = "143",
    year = "2021"
}

@article{Beneke:2009rj,
    author = "Beneke, M. and Falgari, P. and Schwinn, C.",
    title = "{Soft radiation in heavy-particle pair production: All-order colour structure and two-loop anomalous dimension}",
    eprint = "0907.1443",
    archivePrefix = "arXiv",
    primaryClass = "hep-ph",
    reportNumber = "PITHA-09-16, IPPP-09-48, DCPT-09-96, SFB-CPP-09-59",
    doi = "10.1016/j.nuclphysb.2009.11.004",
    journal = "Nucl. Phys. B",
    volume = "828",
    pages = "69--101",
    year = "2010"
}

@article{Czakon:2009zw,
    author = "Czakon, Michal and Mitov, Alexander and Sterman, George F.",
    title = "{Threshold Resummation for Top-Pair Hadroproduction to Next-to-Next-to-Leading Log}",
    eprint = "0907.1790",
    archivePrefix = "arXiv",
    primaryClass = "hep-ph",
    reportNumber = "PITHA-09-17, ITP-SB-09-20",
    doi = "10.1103/PhysRevD.80.074017",
    journal = "Phys. Rev. D",
    volume = "80",
    pages = "074017",
    year = "2009"
}

@article{Ferroglia:2009ep,
    author = "Ferroglia, Andrea and Neubert, Matthias and Pecjak, Ben D. and Yang, Li Lin",
    title = "{Two-loop divergences of scattering amplitudes with massive partons}",
    eprint = "0907.4791",
    archivePrefix = "arXiv",
    primaryClass = "hep-ph",
    reportNumber = "MZ-TH-09-25",
    doi = "10.1103/PhysRevLett.103.201601",
    journal = "Phys. Rev. Lett.",
    volume = "103",
    pages = "201601",
    year = "2009"
}

@article{Ferroglia:2009ii,
    author = "Ferroglia, Andrea and Neubert, Matthias and Pecjak, Ben D. and Yang, Li Lin",
    title = "{Two-loop divergences of massive scattering amplitudes in non-abelian gauge theories}",
    eprint = "0908.3676",
    archivePrefix = "arXiv",
    primaryClass = "hep-ph",
    reportNumber = "MZ-TH-09-28",
    doi = "10.1088/1126-6708/2009/11/062",
    journal = "JHEP",
    volume = "11",
    pages = "062",
    year = "2009"
}

@article{Ahrens:2010zv,
    author = "Ahrens, Valentin and Ferroglia, Andrea and Neubert, Matthias and Pecjak, Ben D. and Yang, Li Lin",
    title = "{Renormalization-Group Improved Predictions for Top-Quark Pair Production at Hadron Colliders}",
    eprint = "1003.5827",
    archivePrefix = "arXiv",
    primaryClass = "hep-ph",
    doi = "10.1007/JHEP09(2010)097",
    journal = "JHEP",
    volume = "09",
    pages = "097",
    year = "2010"
}

\end{document}